\documentclass[journal=jacsat,manuscript=article]{achemso}

\usepackage[version=3]{mhchem} 
\usepackage{color}

\author{Zhedong Zhang}
\affiliation{Department of Physics and Astronomy, SUNY Stony Brook, Stony Brook, NY 11794, USA}
\author{Jin Wang}
\affiliation{Department of Physics and Astronomy, SUNY Stony Brook, Stony Brook, NY 11794, USA}
\alsoaffiliation{Department of Chemistry, SUNY Stony Brook, Stony Brook, NY 11794, USA}
\alsoaffiliation{State Key Laboratory of Electroanalytical Chemistry, Changchun Institute of
Applied Chemistry, Chinese Academy of Sciences, Changchun, Jilin 130022,
P. R. China}
\email{jin.wang.1@stonybrook.edu}
\providecommand{\keywords}[1]{\textbf{\textit{Key words---}} #1}

\title{The Assistance of Molecular Vibrations on Coherent Energy Transfer in Photosynthesis from the View of Quantum Heat Engine}


\begin{document}
\begin{tocentry}
\includegraphics[scale=0.69]{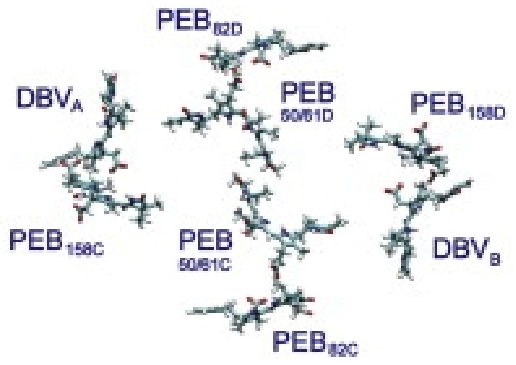}
\includegraphics[width=2.1in,height=1.1in]{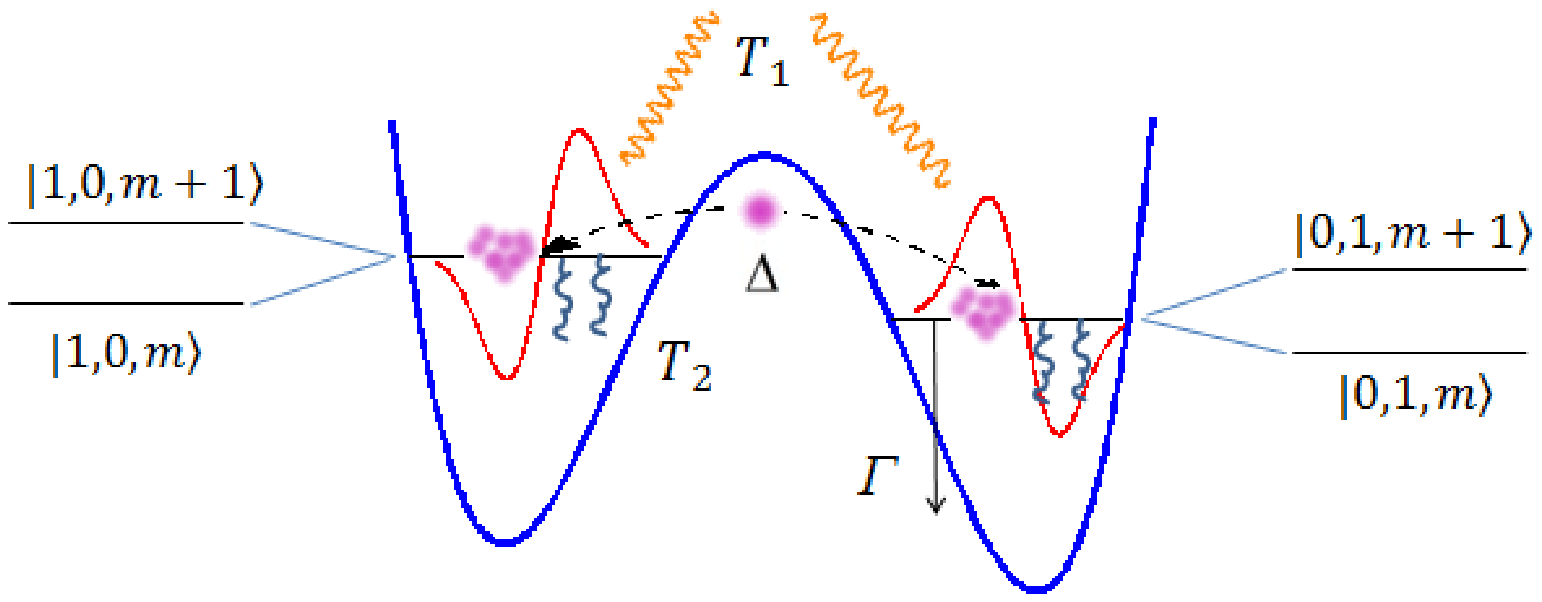}
(left) Illustration of cryptophyte antennae phycoerythrin 545 (PE545) from Ref.\cite{Grondelle10}; (right) our model for PEB$_{50c}$-PEB$_{50d}$ dimer in PE545
\end{tocentry}


\begin{abstract}
Recently the quantum nature in the energy transport in solar cell and light-harvesting complexes have attracted much attention, as being triggered by the experimental observations. We model the light-harvesting complex (i.e., PEB$_{50}$ dimer) as a quantum heat engine (QHE) and study the effect of the undamped intra-molecule vibrational modes on the coherent energy transfer process and quantum transport. We find that the exciton-vibration interaction has {\it non-trivial} contribution to the promotion of quantum yield as well as transport properties of the quantum heat engine at steady state, by enhancing the quantum coherence quantified by entanglement entropy. The perfect quantum yield over 90\% has been obtained, with the exciton-vibration coupling. We attribute these improvements to the renormalization of the electronic couplings effectively induced by exciton-vibration interaction and the subsequent delocalization of excitons. Finally we demonstrate that the thermal relaxation and dephasing can help the excitation energy transfer in PEB$_{50}$ dimer.
\end{abstract}




\section*{Introduction}
Recently the wide-spread interest in uncovering the quantum phenomena in the solar cell and photosynthetic process has been triggered by the experimental investigations of the electronic dynamics in light-harvesting and Fenna-Matthews-Olson (FMO) complexes \cite{Engel07,Engel10,Collini10,Engel12}. The transport of excitation energy absorbed by antenna towards the reaction center (RC) occurs with a perfect efficiency over 90\% \cite{Sauer79,Grondelle06}, which was shown to be strongly correlated to the long-lived quantum coherence between different molecules \cite{Parson07,Fleming09}.

Numerous research made it clear that the exciton energy transport in photosynthetic organism critically depends on the exciton-phonon interaction \cite{Plenio08,Lloyd09,Lloyd08,Fleming12}, besides the electronic coupling between molecules. The exciton-phonon interactions have two types: high-frequency modes from the nuclei vibrations in molecules, and low-frequency modes induced by environmental fluctuations. Actually these two kinds of exciton-phonon interactions in these complexes are associated with low-energy fluctuations of protein immersed in the solvent and intra-molecular vibrations \cite{Renger01,Scholes14}, respectively. Since the intra-molecular vibrations are undamped, it is evidently shown to have significant influence on the coherent energy transfer when energy quanta of vibrational modes is in resonance with the energy splitting of excitons \cite{Reilly14,Moran11,Mancal12}. 

In this work, we study the effect of intra-molecule vibrations on energy transfer processes in the pairs of chromophores, which describes several light-harvesting antennae in the nature. Two important examples are the central $\textup{PEB}_{50c}-\textup{PEB}_{50d}$ dimer in the cryptophyte antennae PE545 (Phycoerythrin545) and the $\textup{Chl}_{b601}-\textup{Chl}_{a602}$ pair in the light-harvesting complex II (LHCII). The natural photosynthetic organism functions in the presence of both solar radiation and low-energy fluctuations of protein. 
In our model, there are three different types of energies involved, namely solar radiation, high-frequency modes from the nuclei vibrations and low-frequency modes from the environmental fluctuations. According to the quantum thermodynamics, QHE converts hot thermal radiation into low-entropy useful work \cite{Mukamel13,Scully97,Breuer02}. We will include the dynamics of discrete vibrational modes between intra-molecules, together with the dynamics of system, due to the comparable relaxation time \cite{Fleming09,fleming10,Schroder06}. We will find that the interaction with vibrational modes has a {\it non-trivial} contribution to the enhancement of the nonequilibriumness flux \cite{Wang08,Zhedong14} trapped by RC and energy transfer efficiency (ETE), by improving the coherence. 

\begin{figure}
\centering
$\begin{array}{c}
 \includegraphics[scale=0.55]{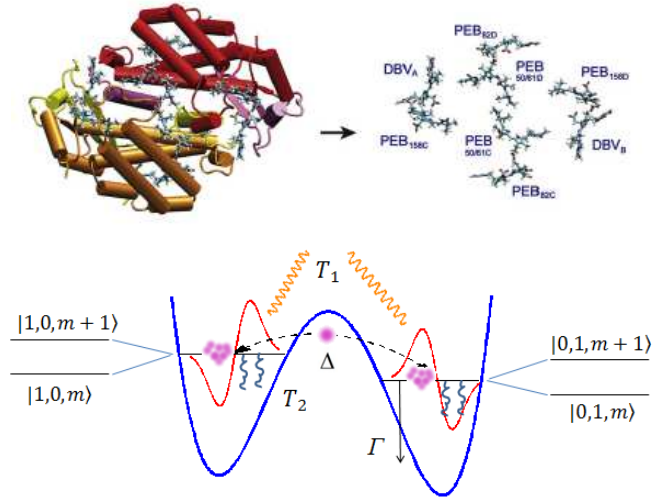}\\
 \includegraphics[scale=0.55]{schematic.eps}\\
 \end{array}$
\caption{(Color online) (top) Illustration of cryptophyte antennae phycoerythrin 545 (PE545); (bottom) Schematic of our model for PEB$_{50c}$-PEB$_{50d}$ dimer in PE545. Standard parameters are taken from Ref.\cite{Grondelle10,Doust04}: $\omega=800\ \textup{cm}^{-1}$ being quasi-resonant with the intra-molecular mode of frequency, $\varepsilon_1-\varepsilon_2=1042\ \textup{cm}^{-1}$, $\Delta=92\ \textup{cm}^{-1}$ and $T_2=300$K.}
\label{Fig}
\end{figure}

\section*{Polaron transform and quantum master equation}
We consider a prototype dimer where each chromophore has an excited state described by exciton with energy $\varepsilon_i$ strongly coupled to a quantized vibrational mode of frequency $\omega$, with the identical coupling strength $\lambda$ for $i=1,2$. The Hamiltonian of dimer and vibrational modes read
\begin{equation}
\begin{split}
H_{ex}=E_0|0\rangle\langle 0|+\varepsilon_1 c_1^{\dagger}c_1+\varepsilon_2 c_2^{\dagger}c_2+\Delta(c_1^{\dagger}c_2+c_2^{\dagger}c_1)
\end{split}
\label{1}
\end{equation}
and $H_{vib}=\hbar\omega(b_1^{\dagger}b_1+b_2^{\dagger}b_2)$. The interaction term is
\begin{equation}
\begin{split}
H_{ex-vib}=\sqrt{2}\lambda\hbar\omega c_1^{\dagger}c_1(b_1+b_1^{\dagger})+\sqrt{2}\lambda\hbar\omega c_2^{\dagger}c_2(b_2+b_2^{\dagger})
\end{split}
\label{2}
\end{equation}
In what follows we will reduce the degrees of the exciton-vibration dynamics, by introducing the correlated (+) and anti-correlated (-) vibrational coordinates: $b_{\pm} = \left(b_1\pm b_2\right)/\sqrt{2}$,
which rewrites the Hamiltonian as 
\begin{equation}
\begin{split}
H_{ex-vib}=\lambda\hbar\omega\left(c_1^{\dagger}c_1-c_2^{\dagger}c_2\right)\left(b_-+b_-^{\dagger}\right)
+\lambda\hbar\omega\left(c_1^{\dagger}c_1+c_2^{\dagger}c_2\right)\left(b_++b_+^{\dagger}\right)
\end{split}
\label{4}
\end{equation}
where the $1^{\textup{st}}$ term in $H_{ex-vib}$ describes the vibration mode with phase difference of $\pi$ while the $2^{\textup{nd}}$ term describes the center-of-mass motion of vibrational modes. Thus the anti-correlated vibrational mode is excited by optical phonons \cite{Madelung96}. Next we will employ the polaron transformation \cite{Nazir09} to eliminate the center-of-mass motion of intra-molecule vibrations and uncover the effect of the exciton-vibration interaction. The generating function can be written as $S = \lambda[(c_1^{\dagger}c_1 -c_2^{\dagger}c_2)(b_-^{\dagger}-b_-)+(c_1^{\dagger}c_1+c_2^{\dagger}c_2)(b_+^{\dagger}-b_+)]$, 
which is obviously anti-Hermitian, namely, $S^{\dagger}=-S$, to ensure the unitarity of the transform. The new Hamiltonian in the polaron frame is obtained by evaluating $e^S H e^{-S}$, and it takes the form of 
\begin{equation}
\begin{split}
\tilde{H} =& \left(\varepsilon_1-2\lambda^2\hbar\omega\right) c_1^{\dagger}c_1
+\left(\varepsilon_2-2\lambda^2\hbar\omega\right)c_2^{\dagger}c_2\\[0.15cm]
& \qquad +\Delta\left[e^{2\lambda(b_-^{\dagger}-b_-)}c_1^{\dagger}c_2+e^{-2\lambda(b_-^{\dagger}-b_-)}c_2^{\dagger}c_1\right]+H_{vib}
\end{split}
\label{6}
\end{equation}
where $H_{vib}=\hbar\omega(b_+^{\dagger}b_++b_-^{\dagger}b_-)$. From Eq.(\ref{6}) it is clear that the center-of-mass motion is independent of the dynamics of exciton so that we can ignore its effect. The exciton-vibration interaction causes a renormalization of the electronic coupling, which as will be shown later, will critically affect the quantum transport. 
Notice that the one-particle approixmation for exciton is not used here and after.

The Hamiltonian in polaron frame provides a picture that the transition of exciton in the dimer is assisted by exchanging energy with the vibrational modes. Physically we can only consider the single-quanta process of vibrations, owing to: ($i$) the low probability of multi-excitation and ($ii$) the energy scale of the exciton-vibration interaction being in quasi-resonance with the gap between the adjacent levels of vibrational modes. Hence the dynamics of intra-molecule vibrations can be restricted into the space spanned by $\{|m\rangle,\ |m+1\rangle\}$, as shown in details in supplementary information (SI). Moreover the occupation $m$ here is the average of particle number in its eigenmodes rather than the thermal occupation of bosons. The total Hilbert space is $\mathcal{H}=\mathcal{H}_{\textup{elec}}^{(1)}\otimes\mathcal{H}_{\textup{elec}}^{(2)}\otimes\mathcal{H}_{\textup{vib}}$.

In light-harvesting complexes, i.e., PEB$_{50}$ dimer and LHCII pair (or FMO complex), the excitons need to interact with radiation environment from solar as well as low-energy fluctuations of protein immersed in the solvent, in order to realize the energy transfer to RC. In addition, we need a connection of one site in the dimer (labelled by 2 in our notation) to RC, described by a trapping rate $\Gamma$, in order to generate the output work. The interactions to radiation and protein reservoirs read (notice that this is not influenced by the polaron transform introduced before, since the polaron transform only operates in the subspace of the system Hamiltonian)
\begin{equation}
\begin{split}
H_{\textup{int}}= \sum_{i=1}^2\sum_{\textbf{k},p}g_{\textbf{k}p}
\left(c_i+c_i^{\dagger}\right)
\left(a_{\textbf{k}p}+a_{-\textbf{k},p}^{\dagger}\right)
+\sum_{i=1}^2\sum_{\textbf{q},s}\gamma_{\textbf{q},s}f_i c_i^{\dagger}c_i\left(b_{\textbf{q}s}+b_{-\textbf{q},s}^{\dagger}\right)
\end{split}
\label{9}
\end{equation}
where $a_{\textbf{k}p}$ and $b_{\textbf{q}s}$ are the bosonic operators for radiation and low-energy fluctuation reservoirs, respectively. $p$ and $s$ denote the polarizations of the modes of radiation and low-energy fluctuation reservoirs, respectively.  
Thus the total Hamiltonian of the system and environments reads $H = \tilde{H} + H_{\textup{bath}} + H_{\textup{int}}$, where $H_{\textup{bath}} = \sum_{\textbf{k},p}a_{\textbf{k}p}^{\dagger}a_{\textbf{k}p} + \sum_{\textbf{q},s}b_{\textbf{q}s}^{\dagger}b_{\textbf{q}s}$. 
Based on the perturbation theory, the whole solution to density operator can be written as $\rho_{SR}=\rho_S(t)\otimes\rho_R(0)+\rho_{\delta}(t)$ with the traceless term in higher orders of coupling. Since the time scale associated with the environmental correlations is much smaller than the time scale of the system over which the state varies appreciably, the quantum master equation (QME) for reduced density matrix of systems can be derived under the so-called Markoff approximation
\begin{equation}
\begin{split}
& \frac{d\rho}{dt}=\frac{i}{\hbar}[\rho,\tilde{H}]+\frac{1}{\hbar^2}\sum_{a,b=0}^7\sum_{c,d=0}^7 T_{ad,bc}\Big(|a\rangle\langle b|\rho|c\rangle\langle d|-\delta_{ad}|c\rangle\langle b|\rho\Big)\\[0.1cm]
& \qquad\quad +\frac{1}{\hbar^2}\sum_{a=0}^7\sum_{b=0}^7\sum_{c=0}^7 R_{abc}\Big(|a\rangle\langle b|\rho|c\rangle\langle c| -\delta_{ac}|c\rangle\langle b|\rho\Big)+\textup{h.c.}+\Gamma\mathcal{D}_{tr}(\rho)
\end{split}
\label{10}
\end{equation}
where the decay rate $T_{ad,bc}/\hbar^2 = v_{dc}\sum_{k,l}\sum_{m,n=0}^7\Gamma_{kl}^{T_1}v_{mn}U_{km}^{\dagger}U_{ak}U_{nl}U_{lb}^{\dagger}$ and the dephasing rate $R_{abc}/\hbar^2 = \sum_{i=0}^7\sum_{k,l}\gamma_{kl}^{T_2}P_iP_c U_{ik}U_{li}^{\dagger}U_{ka}^{\dagger}U_{bl}$. The superoperator describing the trapping by RC is 
$\mathcal{D}_{tr}(\rho) = \sum_{n=m}^{m+1}
(|0,0,n\rangle\langle 0,1,n|\rho|0,1,n\rangle\langle 0,0,n|-|0,1,n\rangle\langle 0,1,n|\rho|0,1,n\rangle\langle 0,1,n|)$. $\Gamma_{kl}^{T_1}=\int \textup{d}\textbf{q}/(2\pi)^2 g_{\textbf{q}}^2(n_{\textbf{q}}+1)\delta(\omega'_{kl}-\nu_{\textbf{q}})$
for $k>l$, or $\int \textup{d}\textbf{q}/(2\pi)^2 g_{\textbf{q}}^2 n_{\textbf{q}}\delta(\omega'_{kl}+\nu_{\textbf{q}})$ for $k<l$. For the environment of protein in the solvent, we use the Debye spectral density: $J(\omega)=\frac{2E_R}{\pi\hbar}\frac{\omega\omega_d}{\omega^2+\omega_d^2}$ where $E_R$ is so-called reorganization energy, and $\gamma_{kl}^{T_2}=J(|{\omega'}_{kl}|)n(|{\omega'}_{kl}|)$ for $k>l$, $J(|{\omega'}_{kl}|)\left[n(|{\omega'}_{kl}|)+1\right]$ for $k<l$, or $\frac{2E_R}{\pi\hbar}\frac{k_B T_2}{\hbar\omega_d}$ for $k=l$. $U$ is the unitary matrix that diagonalizes the Hamiltonian matrix $\tilde{H}$. In the Liouville space, the QME can be formulated as two-component form
\begin{equation}
\begin{split}
\frac{\partial}{\partial t}\begin{pmatrix}
                            \rho_p\\[0.12cm]
                            \rho_c\\
                           \end{pmatrix}
                          =\begin{pmatrix}
                            \mathcal{M}_p & \mathcal{M}_{pc}\\[0.12cm]
                            \mathcal{M}_{cp} & \mathcal{M}_c\\
                           \end{pmatrix}
                           \begin{pmatrix}
                            \rho_p\\[0.12cm]
                            \rho_c\\
                           \end{pmatrix}
\end{split}
\label{11}
\end{equation}
Here, $\rho_p$ and $\rho_c$ represent the population and coherence components of density matrix, respectively. To obtain the population dynamics, we need to project the QME into population space, by eliminating the coherence components using Laplace transform \cite{Zhedong14}. Based on the view point of QHE, the whole system should work at steady state, which is what we are interested in this Letter. Thus in the long-time limit, the QME at steady state reads
\begin{equation}
\begin{split}
\left(\mathcal{M}_p-\mathcal{M}_{pc}\mathcal{M}_c^{-1}\mathcal{M}_{cp}\right)\rho_p^{ss}=0
\end{split}
\label{12}
\end{equation}
where we introduce $\mathcal{A}\equiv\mathcal{M}_p-\mathcal{M}_{pc}\mathcal{M}_c^{-1}\mathcal{M}_{cp}$ and can define the transfer matrix as $T_{mn}=\mathcal{A}_{nm}\rho_{m}^p$. Thus the net-nonequilibrium-flux is $c_{mn}=\mathcal{A}_{nm}\rho_m^p-\textup{min}(\mathcal{A}_{nm}\rho_m^p,\mathcal{A}_{mn}\rho_n^p)$, which can be decomposed into the superposition of several closed loops \cite{Zhedong14,Qian82}. Actually $c_{mn}$ quantifies the detailed-balance-breaking and time-irreversibility. In reality, people are principally able to observe the flux trapped by RC, which is defined as $\mathcal{J}_{trap}=\Gamma(\rho_{4}^p+\rho_5^p)$. This is consistent with the definition of $c_{mn}$ since the transition induced by RC is unidirectional. As we will see later, $\mathcal{J}_{trap}$ will play an significant role in the discussion of transport properties of this QHE.  

\section{Coherent energy transfer and quantum transport}
\subsection{Energy transfer efficiency and flux trapped by RC}
\begin{figure}
\centering
$\begin{array}{ccc}
 \includegraphics[scale=0.36]{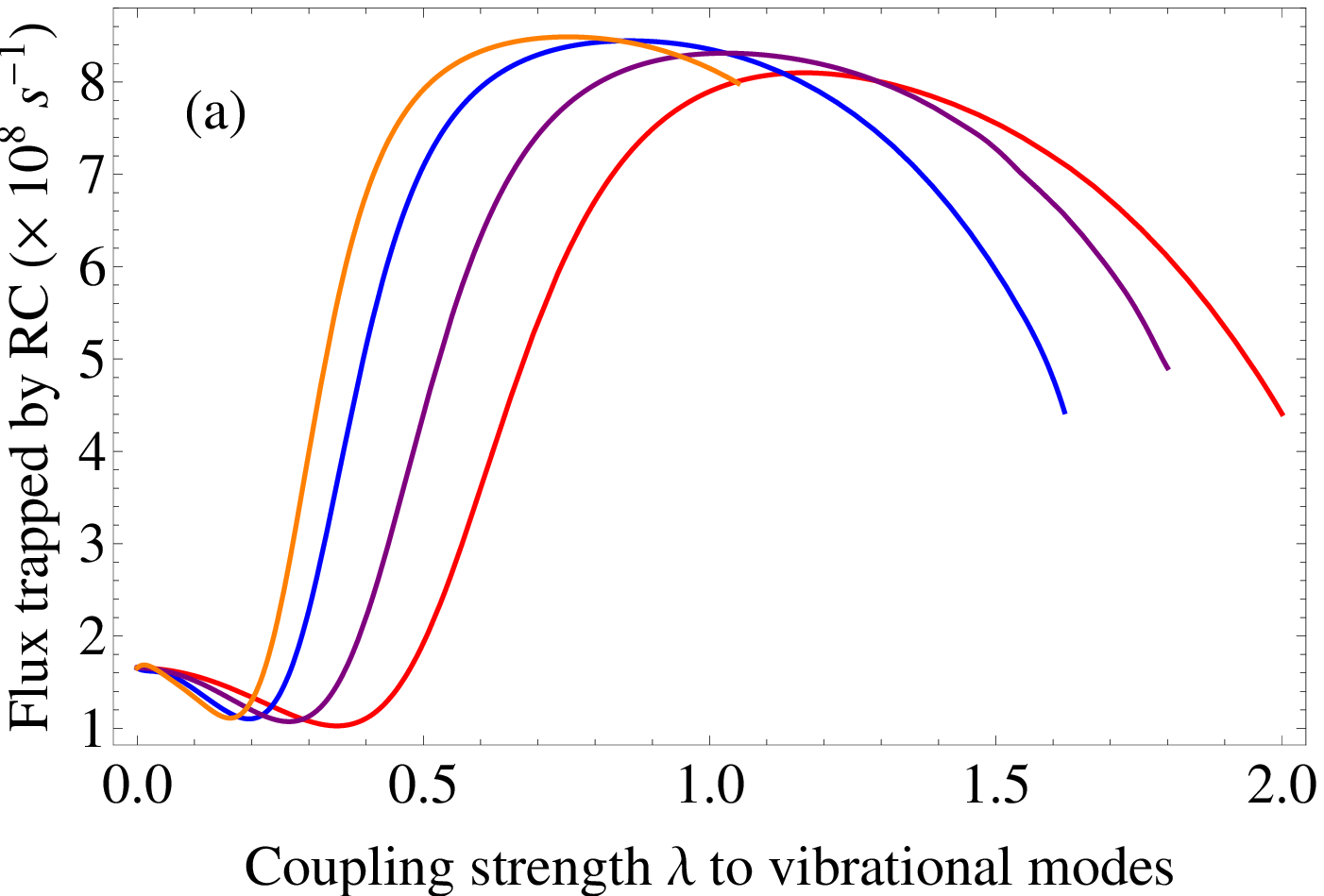}
&\includegraphics[scale=0.39]{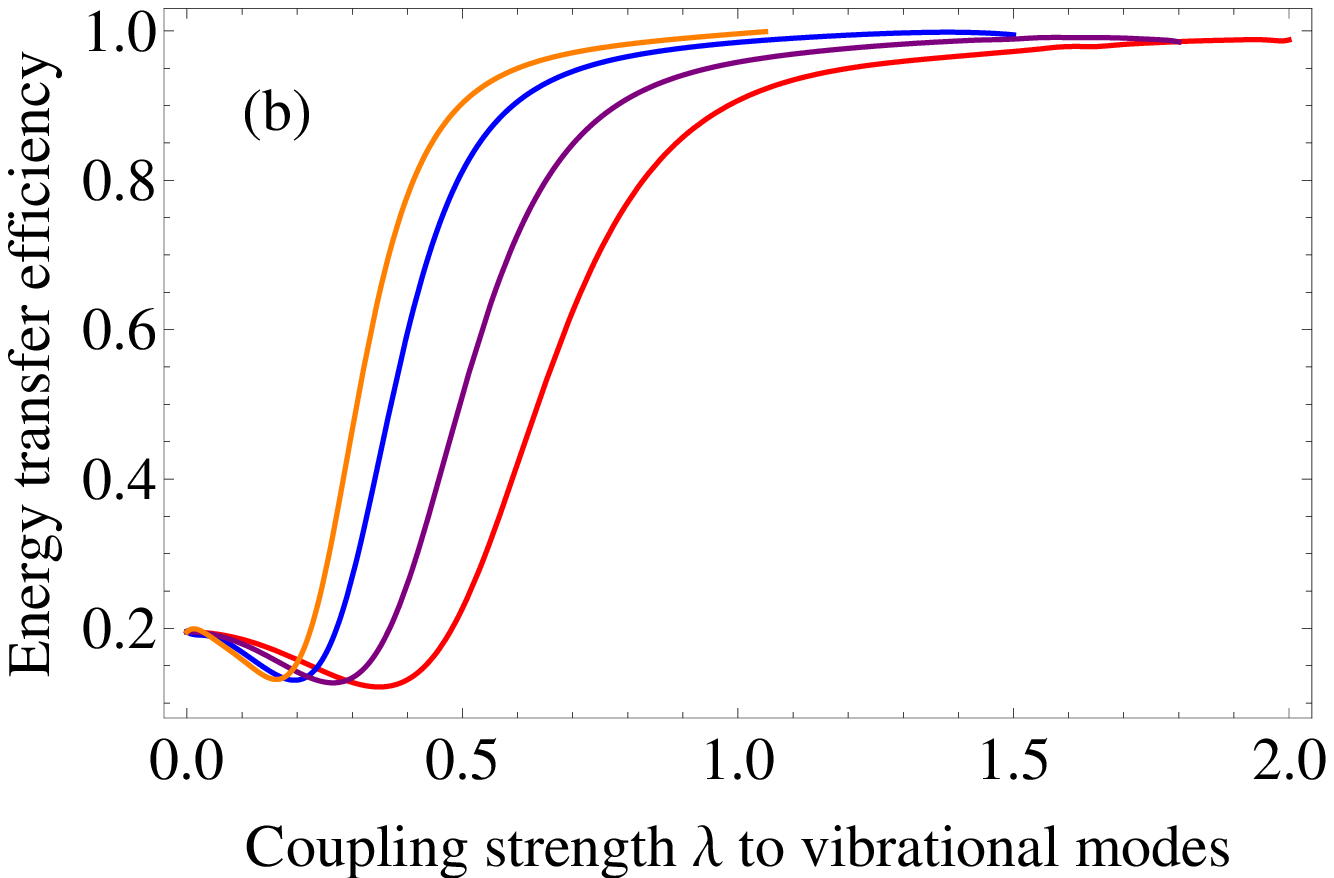}
&\includegraphics[scale=0.367]{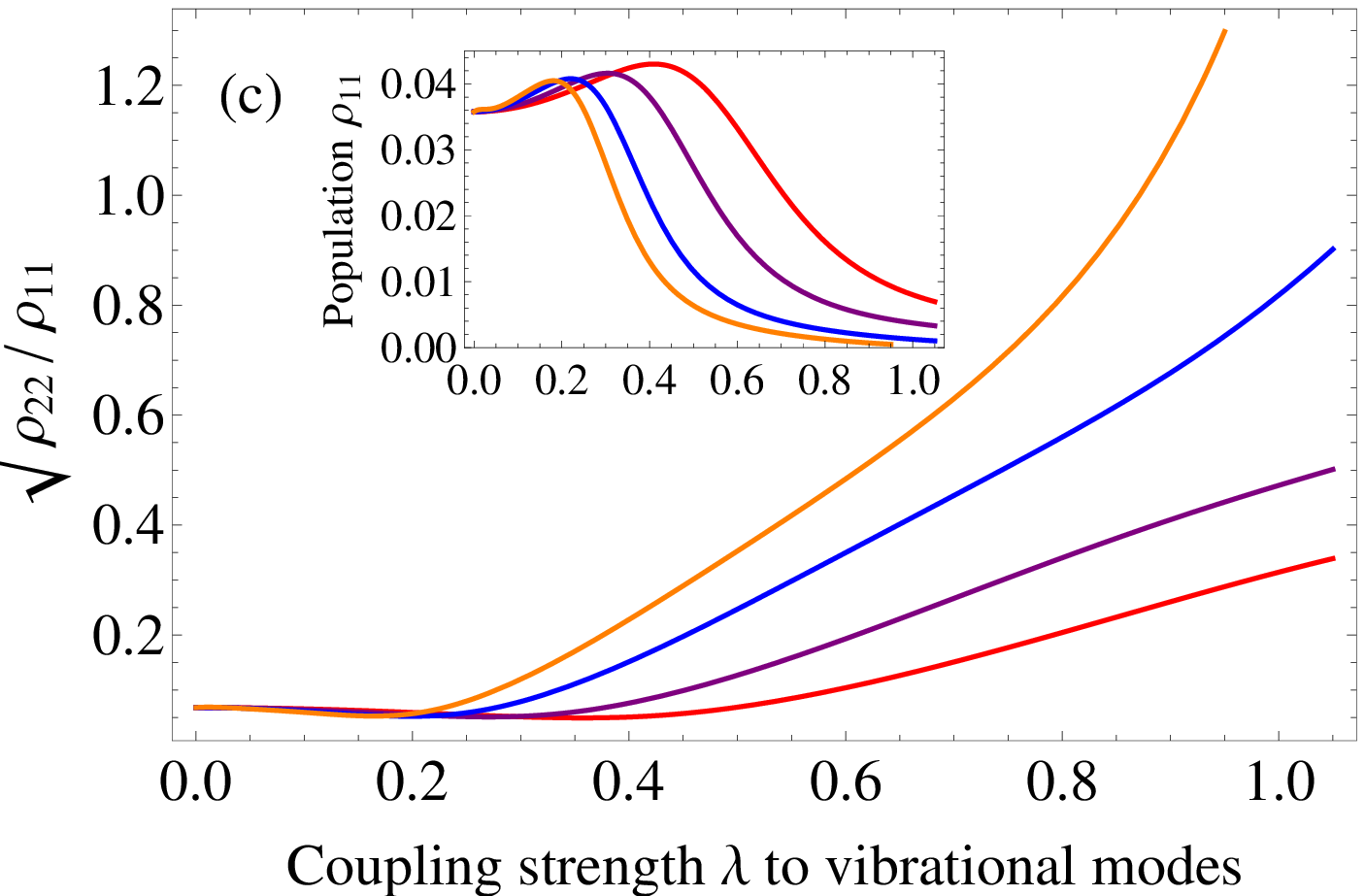}\\
 \includegraphics[scale=0.36]{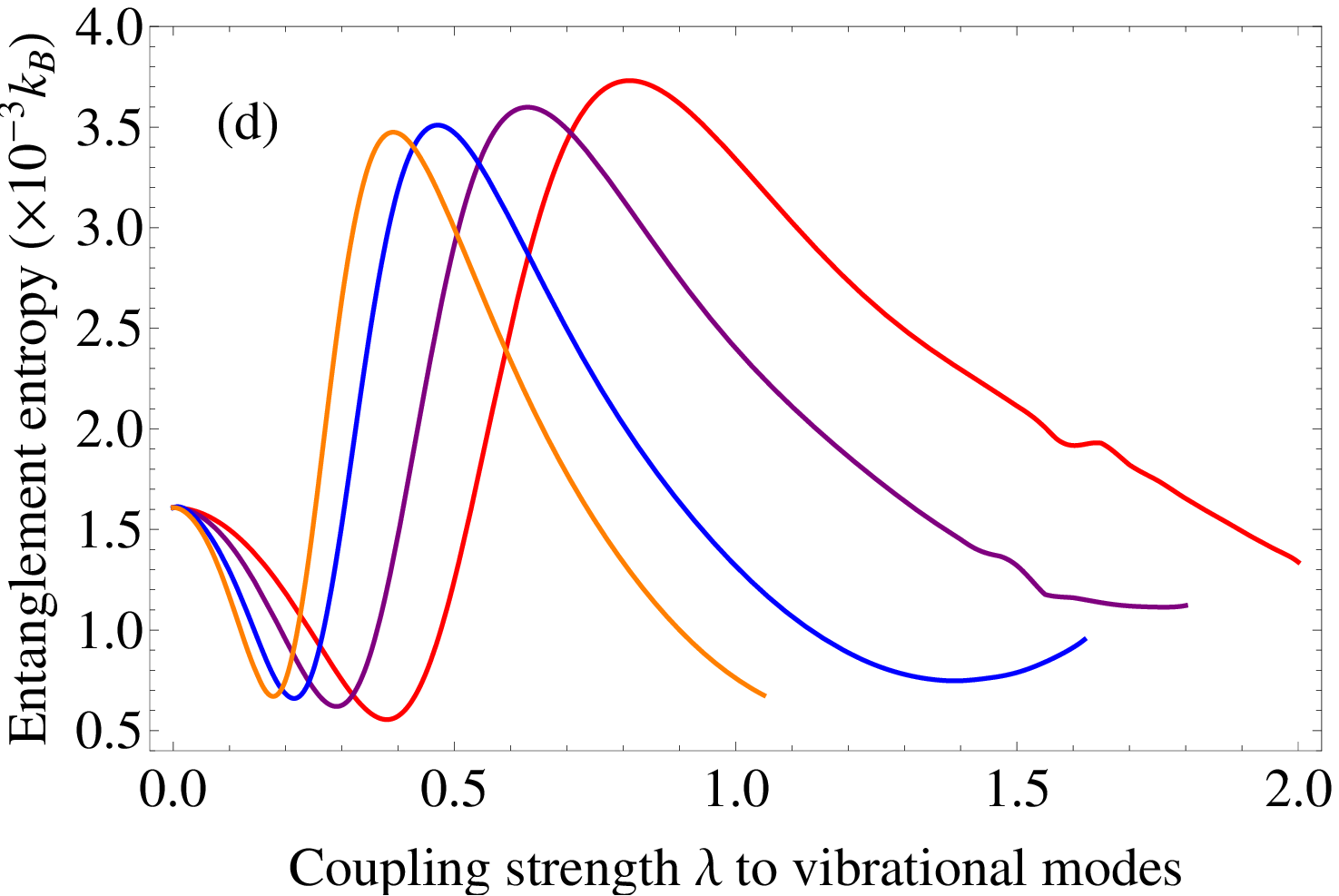}
&\includegraphics[scale=0.378]{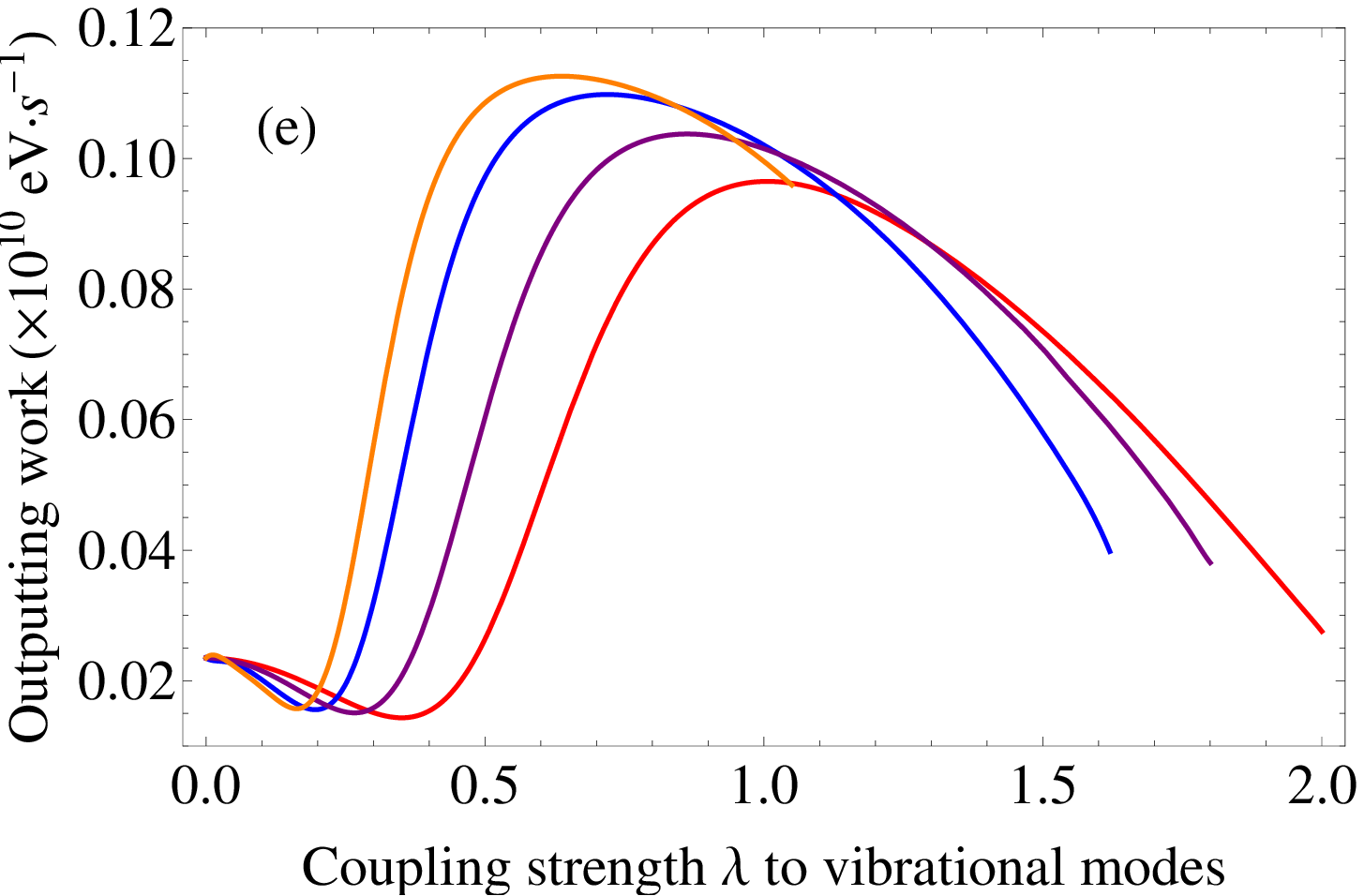}
&\includegraphics[scale=0.349]{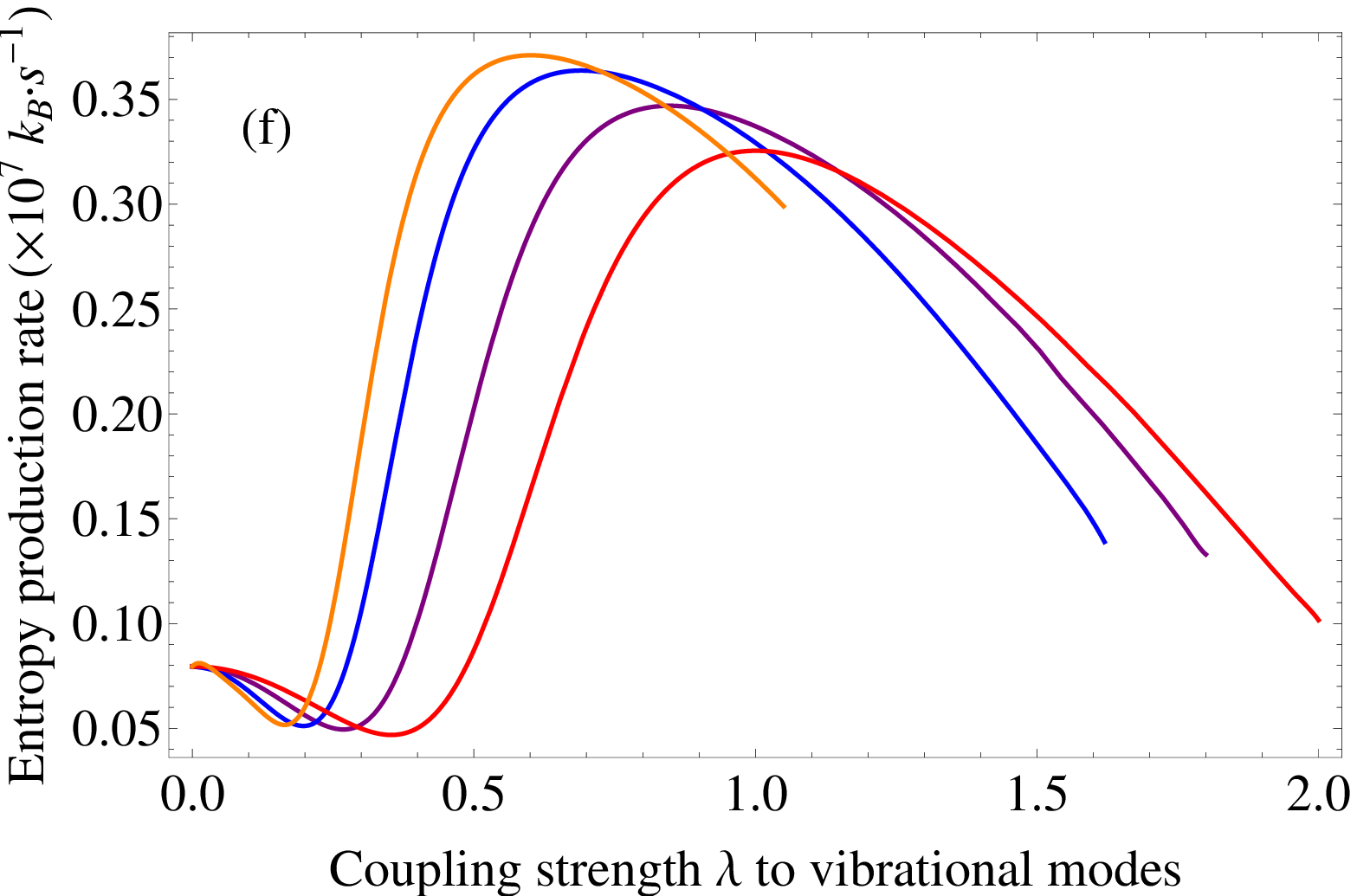}\\
 \end{array}$
\caption{(Color online) (a) Flux trapped by RC, (b) energy transfer efficiency, (c) the population on site 1 (small figure) \& the square root of ratio of the populations of two chromophores in the PE545 dimer (large figure), (d) the entanglement entropy (which also quantifies the coherence effect), (e) work generated by QHE to RC and (f) entropy production (EPR) vary as a function of coupling strength of exciton to intra-molecule vibrational modes. Red, purple, blue and orange lines correspond to $m=0, 1, 3, 5$, respectively, where $m$ represents the bosonic occupation of the vibrational modes. Standard paramenters are $\omega=800$cm$^{-1}$, $\epsilon_1-\epsilon_2=1042$cm$^{-1}$, $\Delta=92$cm$^{-1}$ and $T_2=300$K \cite{Grondelle10,Doust04}.}
\label{Fig.1}
\end{figure}

Now we are able to discuss the coherent energy transfer in the dimer, after the absorption of photons from the Sun. In the natural light-harvesting complexes, the reorganization energy is $E_R=34\textup{cm}^{-1}$, the cut-off frequency is $\omega_d=50\ \textup{fs}^{-1}$, the temperature of the radiation reservoir is $T_1=5780$K \cite{Kassal13} and the trapping rate by RC is set to be $\Gamma=1\textup{ps}^{-1}$. In PEB$_{50}$ dimer the dipole moments in Eq.(\ref{9}) are $f_1=1.0,\ f_2=-0.9$ \cite{Grondelle10}. First we introduce the energy transfer efficiency (ETE) \cite{Lloyd09,Lloyd08}
\begin{equation}
\begin{split}
\eta = \frac{\mathcal{J}_{trap}}{\mathcal{J}_{trap}+K_{de}}
\end{split}
\end{equation}
where the decay rate is $K_{de}=\sum_{\mu=0}^1\sum_{\nu=2}^5\mathcal{A}_{\mu\nu}\rho_{\nu}^p - \Gamma(\rho_4^p+\rho_5^p)$, which quantifies the probability going back to ground state per unit time. Fig.{\ref{Fig.1}}(a) and {\ref{Fig.1}}(b) illustrates the effect of coupling strength of exciton to intra-molecule vibrations on the trapping flux by RC as well as ETE. As we can see, both of the flux and ETE show a sharp increase as the coupling strength becomes large, besides a small decrease at beginning. The reason for such decrease is the corresponding small increase of population on site 1 as shown in Fig.\ref{Fig.1}(c), which indicates that the exciton becomes more localized at very weak coupling to molecular vibrations. Significantly, the large coupling to vibrations leads to the optimization of ETE of the system, namely, over 90\%. Physically, we can understand it as follows: the exchange of the energy between excitons and vibrational modes leads to the renormalization of the electronic coupling, which effectively amplifies the magnitude of the electronic couplings by a factor of $\sqrt{m+1}$ as shown in the Hamiltonian. Consequently, the exciton transport is accelerated. 
On the other hand, Fig.\ref{Fig.1}(a) also indicates that the nonequilibriumness of the system can be enhanced by the exciton-vibration dynamics, as quantified by the trapping flux. In the regime of strong exciton-vibration coupling, namely, $\lambda\gg 1$, the states become mainly vibrational and the excitonic transport is suppressed, as relfected by the decay of the flux illustrated in Fig.\ref{Fig.1}(a), though the ETE still saturates to a perfect value of 100\%. In this sense, we know that the energy transfer efficiency is necessary, but not sufficient for the description of the excitation energy transport in the photosynthesis antenna. In addition to this, other quantities, such as output work (heat current into RC) and EPR, are essential as well to completely describe the quality of this QHE on excitation energy transport.

Moreover, by comparison of fluxes and ETE for different occupations of vibrations, it is clear that the higher excited mode of intra-molecule vibrations is more efficient for the enhancement of ETE and flux trapped by RC. However, the environment around the chromophores always satisfies the condition $\hbar\omega\sim k_B T_2$, thus only those lowest vibrational modes can be populated. Hence we show the results for $m=0,1,3,5$ here. Particularly, the ETE and flux for $m=3$ are of 72\% and 75\% improvements with respect to $m=0$, at $\lambda=0.6$.

The energy transfer between the PEB$_{50}$ dimer and RC acquires the delocalization of the excitons. We now investigate the trends of delocalization under the influence of exciton-vibration interaction. As is shown in Fig.\ref{Fig.1}(c), population of high-lying exciton state has a rapid, non-exponential decay as the coupling to vibrational modes increases, which can be traced back to the coherent transition from $|1,0,m\rangle$ to $|0,1,m+1\rangle$ and from $|1,0,m+1\rangle$ to $|0,1,m\rangle$. On the other hand, the delocalization of the wave packet is demonstrated in the large figure in Fig.\ref{Fig.1}(c), by increasing the coupling strength to vibrations. In fact, the coupling between excitons and vibrational modes leads to an effective amplification of electronic coupling by a factor of $\sqrt{m+1}$ as mentioned before, between the two molecules in a dimer. This is particularly reflected in the matrix elements: $\langle 1,0,m|\tilde{H}|0,1,m+1\rangle,\ \langle 1,0,m|\tilde{H}|0,1,m\rangle$ and $\langle 1,0,m+1|\tilde{H}|0,1,m\rangle$.

\subsection{Coherence effect}
Since the dynamics of intra-molecule vibrations is considered together with excitons, we need to study the influence of exciton-vibration interaction on the coherence effect. Notice that the coherence effect mentioned here refers to the entanglement entropy and the sum of the off-diagonal elements of density matrix with different electronic states (quantum coherence). Due to the strong correltaion between entanglement and coherence, we use the quantum entanglement to quantify the coherence effect hereafter and the coherence will be enclosed in SI. To calculate the entanglement entropy, we first need to diagolize the density matrix at steady state $\rho=\sum_{f}P_f|\psi_f\rangle\langle\psi_f|$, where $|\psi_f\rangle = \sum_{f}\mathcal{U}_{fn}|n_1(f),n_2(f),m(f)\rangle$ and $\mathcal{U}$ is the unitary transform matrix in the diagonalization of density matrix. For each component of pure state, the density matrix reads, by partial tracing over the freedoms of other sites except the first one ($n_3\equiv m$)
\begin{equation}
\begin{split}
\rho_f^{(1)} & = \sum_{\nu=0}^7\langle n_2(\nu),m(\nu)|\psi_f\rangle\langle \psi_f|n_2(\nu),m(\nu)\rangle\\[0.1cm]
& =\sum_{j,l=0}^7\left(\mathcal{U}_{jf}\mathcal{U}_{fl}^{\dagger}\sum_{k=0}^7\prod_{s=2}^3\delta_{n_s(k),n_s(j)}\delta_{n_s(k),n_s(l)}\right)|n_1(j)\rangle\langle n_1(l)|
\end{split}
\label{ep}
\end{equation}
whose eigenvalues are $\lambda_{\pm}^f$. Then the entanglement entropy of each pure component is $S_f^{(1)}=-k_B(\lambda_+^f\textup{ln}\lambda_+^f + \lambda_-^f\textup{ln}\lambda_-^f)$, which subsequently gives the total entanglement entropy at steady state
\begin{equation}
\begin{split}
S = -k_B\sum_{f=0}^7P_f\left(\lambda_+^f\textup{ln}\lambda_+^f + \lambda_-^f\textup{ln}\lambda_-^f\right)
\end{split}
\label{ept}
\end{equation}
Actually the formal definition of the entanglement entropy in mixture ensemble is still an open question. The rational for our choice of the expression in Eq.(\ref{ept}) for our system is: the excitonic basis describing the delocalized excitons is the one which gives the unique diagonal form of the steady-state density matrix and these delocalized excitons contain the coherence between the local sites and play an important role in experiments. Moreover, the off-diagonal elements of the Hamiltonian also leads to the entanglement in the excitonic basis. As illustrated by Fig.\ref{Fig.1}(d), 
the exciton-vibration interaction improves the coherence at first, owing to the energy-exchange between excitons and vibrational modes. However, based on the discussion on the delocalization above, we know that the interaction with vibrational modes leads to the enhancement of the delocalization of exciton wave packet. But we can further see from Fig.\ref{Fig.1}(c) that the population on site 1 of the excitonic state at large exciton-vibration coupling is closed to 0. {\it This indicates the suppression of the back-and-forth oscillation between the two localized excitonic states, which consequently causes the downhill trends of the coherence.} On the other hand, the exciton-vibration leads to the Rabi oscillation with the frequency $\sim \sqrt{\delta\omega^2+\lambda^2\omega^2}$, which indicates a rapid oscillation as well as the mismatching between the energy splitting of excitons and quanta of vibrational modes at large coupling strength. Thus the quantum entanglement and coherence will eventually vanish in long time limit. This is partially in contrast to the previous predictions \cite{Mancal12,tiwari12}.
Furthermore, by the comparison between Fig.\ref{Fig.1}(a), \ref{Fig.1}(b) and \ref{Fig.1}(d), it is found that in the range of coupling to vibrations before the extremum of entanglement, the improvement of quantum coherence gives rise to a rapid growth of flux trapped by RC and ETE, namely, $\lambda \simeq 0.39, 0.46, 0.62, 0.8$ for $m=5,3,1,0$ respectively. In other words, the coherent energy transfer can lead to a significant improvement of quantum yield. The incoherent energy transfer becomes important on the further improvement of ETE after the coherent process.

\section{Macroscopic transport}
According to quantum thermodynamics, the observables on macroscopic level serve as an important role in the function of photosynthetic organisms as a QHE. We now investigate two representative quantities: output work and entropy production rate (EPR). The former provides a critical measure of the macroscopic quality of the QHE and the latter quantifies the nonequilibriumness of the system on macroscopic level.
  The $1^{\textup{st}}$ and $2^{\textup{nd}}$ laws in thermodynamics give $\dot{Q}_1-\dot{Q}_2-\dot{Q}_{tr} = \dot{E}$, $\sigma+\dot{S}=\dot{S}_t$, where the output work to RC can be calculated by $\dot{Q}_{tr} = \textup{Tr}[\tilde{H}\mathcal{D}_{trap}(\rho)]$. $\dot{S}$ and $\dot{S}_t$ are the rates of system entropy and EPR, respectively. Due to the assumption of large environments with the negligible back influence from system to environments, the entropy flux flowing from system to environments reads $\sigma=-\frac{\dot{Q}_1}{T_1}+\frac{\dot{Q}_2}{T_2}+\frac{\dot{Q}_{tr}}{T_{RC}}$ where $T_{RC}$ is the temperature of RC. At steady state, no energy consumption occurs inside the system so that $\dot{Q}_{tr}=\dot{Q}_1-\dot{Q}_2,\ \dot{S}_t=\sigma$. In several natural light-harvesting antennae the temperature in RC is always the same as that for low-energy solvated protein fluctuations, namely, around the room temperature. In Fig.\ref{Fig.1}(e) and \ref{Fig.1}(f), the output work and EPR have similar behaviors: the coupling of excitons to vibrational modes promotes the coherent energy transport which leads to a rapid growth of the work and EPR. The incoherent transport caused by higher coupling strength to vibrations further saturates the work and EPR to particular values (distinguished by different levels of intra-molecule vibrations). For the strong exciton-vibration coupling that $\lambda \gg 1$, the incoherent transport leads to the suppression of output work as well as EPR, since the vibration dominates the feature of states.

\section{Effect of thermal relaxation on coherent energy transfer}
\begin{figure}
\centering
$\begin{array}{cc}
 \includegraphics[scale=0.41]{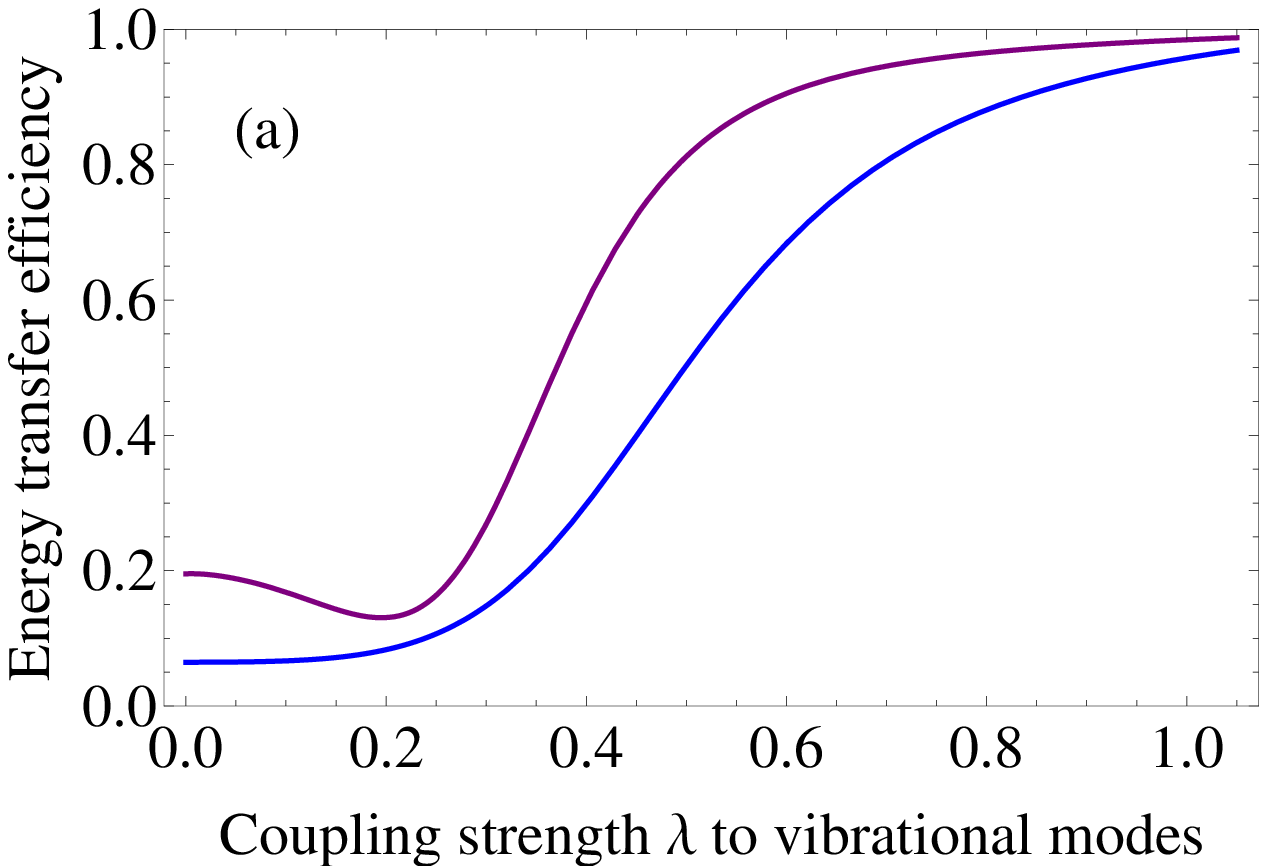}
&\includegraphics[scale=0.405]{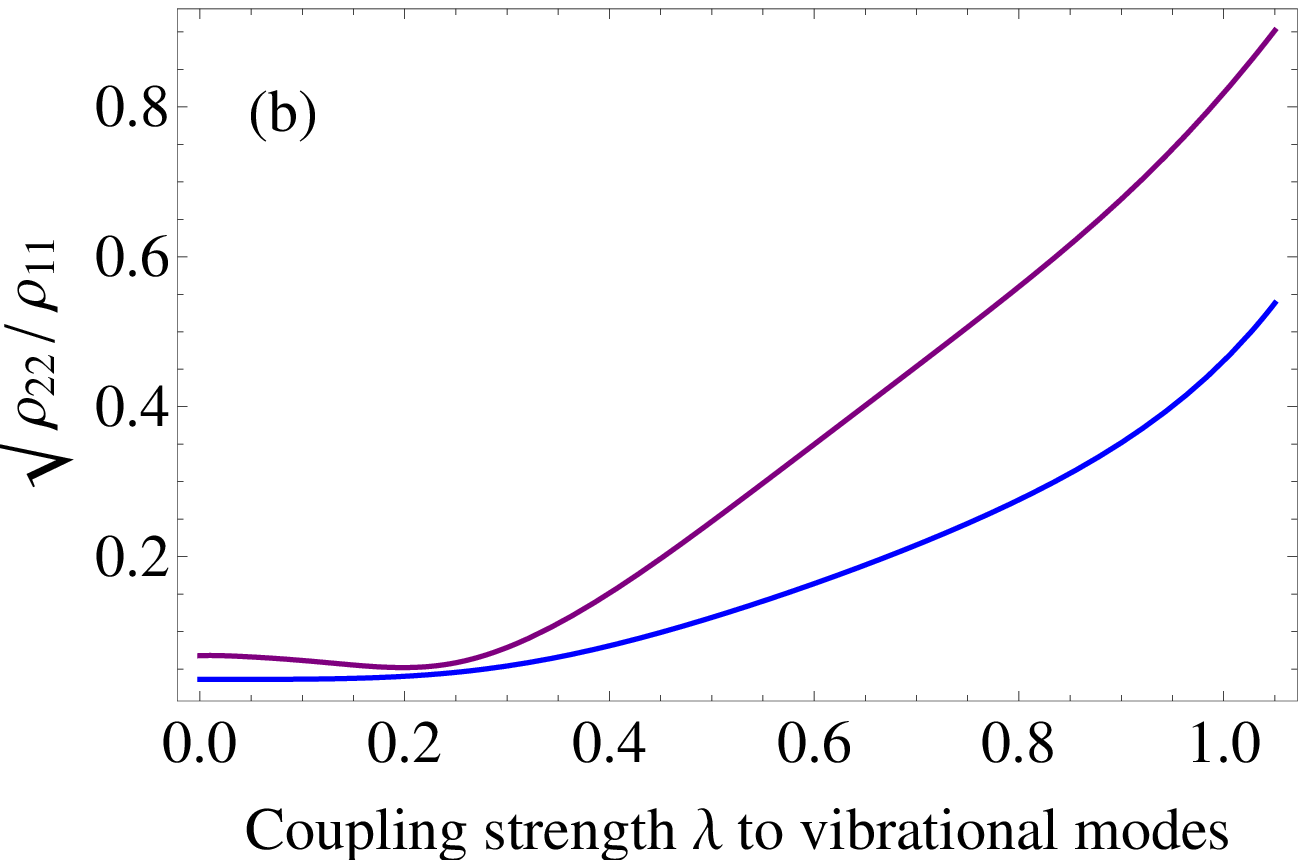}\\
 \end{array}$
\caption{(Color online) Comparison between the regimes of weak and intermediate couplings of exciton-bath, as a function of coupling to vibrational modes. Blue and purple curves correspond to the reorganization energies $E_R=4\textup{cm}^{-1}$ and $34\textup{cm}^{-1}$, respectively.}
\label{Fig.5}
\end{figure}

Now we study the influence of the low-energy thermal bath described by a continuous distribution of harmonic oscillators on the coherent energy transfer in the exciton-vibration dimer. The interaction between exciton and bath is governed by Debye spectral density with the cut-off frequency $\omega_d=50\ \textup{fs}^{-1} < k_B T_2$. Here we consider two regimes of coupling strength between PEB$_{50}$ dimer and bath: very weak coupling $E_R=4\textup{cm}^{-1}$ and intermediate coupling $E_R=34\textup{cm}^{-1}$. As is shown in Fig.\ref{Fig.5}, the dephasing and fluctuation induced by the low-energy protein motion gives rise to the promotion of quantum yield illustrated in Fig.\ref{Fig.5}(a), since the motion of excitons globally becomes more delocalized when the coupling increases, as reflected in Fig.\ref{Fig.5}(b).

\section{Summary and remarks}
In conclusion, we investigated the effect of intra-molecule vibrational modes on the coherent energy transfer in the light-harvesting complexes. It was demonstrated that the exciton-vibration interaction led to a non-trivial improvement of coherent energy transfer by the enhancement of coherence. Furthermore we also show that the exciton-vibration coupling can give rise to a perfect quantum yield (over 90\%) for energy transport, at steady state. 
Our study provides the insights for the exploration of the intre- and intra-molecular vibrations on multi-molecule systems, i.e., the natural light-harvesting complexes LHCII with 32 chromophores and FMO complex with 7 molecules, to see how the exciton-vibration interaction affects the energy transfer pathways.

%

\begin{acknowledgement}
We acknowledge the support from the grant NSF-MCB-0947767
\end{acknowledgement}

\begin{suppinfo}
Details of the derivations of equations are supplied. 
\end{suppinfo}


\begin{thebibliography}{}
\bibitem{Engel07}
Engel, G.\ S.; {\it et al.} Evidence for Wavelike Energy Transfer through Quantum Coherence in Photosynthetic Systems. {\it Nature} \textbf{2007}, {\it 446}, 782-786
\bibitem{Engel10}
Panitchayangkoon, G.; Hayes, D.; Fransted, K.\ A.; Caram, J.\ R.; Harel, E.; Wen, J.; Blankenship, R.\ E.; Engel, G.\ S. Long-lived Quantum Coherence in Photosynthetic Complexes at Physiological Temperature. {\it Proc. Natl. Acad. Sci. USA} \textbf{2010}, {\it 107}, 12766-12770
\bibitem{Collini10}
Collini, E.; Wong, C.\ Y.; Wilk, K.\ E.; Curmi, P.\ M.\ G.; Brumer, P.; Scholes, G.\ D. Coherently Wired Light-Harvesting in Photosynthetic Marine Algae at Ambient Temperature. {\it Nature} \textbf{2010}, {\it 463}, 644-647
\bibitem{Engel12}
Harel, E.; Engel, G.\ S. Quantum Coherence Spectroscopy reveals Complex Dynamics in Bacterial Light-Harvesting Complex 2 (LH2). {\it Proc. Natl. Acad. Sci. USA} \textbf{2012}, {\it 109}, 706-711
\bibitem{Sauer79}
Sauer, K. Photosythesis-The Light Reactions. {\it Annu. Rev. Phys. Chem.} \textbf{1979}, {\it 30}, 155-178
\bibitem{Grondelle06}
van Grondelle, R.; Novoderezhkin, V.\ I. 
Energy Transfer in Photosynthesis: Experimental Insights and Quantitative Models
. {\it Phys. Chem. Chem. Phys.} \textbf{2006}, {\it 8}, 793-807
\bibitem{Parson07}
Parson, W.\ W. Long Live Electronic Coherence! {\it Science} \textbf{2007}, {\it 316}, 1438-1439
\bibitem{Fleming09}
Ishizaki, A.; Fleming, G.\ R. Theoretical Examination of Quantum Coherence in a Photosynthetic System at Physiological Temperature. {\it Proc. Natl. Acad. Sci. USA} \textbf{2009}, {\it 106}, 17255-17260
\bibitem{Plenio08}
Plenio, M.\ B.; Huelga, S.\ F. Dephasing-assisted Transport: Quantum Networks and Biomolecules. {\it New J. Phys.} \textbf{2008}, {\it 10}, 113019-113032
\bibitem{Lloyd09}
Rebentrost, P.; Mohseni, M.; Kassal, I.; Lloyd, S.; Aspuru-Guzik, A. Environment-assisted Quantum Transport. {\it New J. Phys.} \textbf{2009}, {\it 11}, 033003-033014
\bibitem{Lloyd08}
Mohseni, M.; Rebentrost, P.; Lloyd, S.; Aspuru-Guzik, A. Environment-assisted Quantum Walks in Photosynthetic Energy Transfer. {\it J. Chem. Phys.} \textbf{2008}, {\it 129}, 174106-174114
\bibitem{Fleming12}
Ishizaki, A.; Fleming, G.\ R. Quantum Coherence in Photosynthetic Light Harvesting. {\it Annu. Rev. Condens. Matter Phys.} \textbf{2012}, {\it 3}, 333-361
\bibitem{Renger01}
Renger, T.; May, V.; K$\ddot{\textup{o}}$hn, O. Ultrafast Excitation Energy Transfer Dynamics in Photosynthetic Pigment-Protein Complexes. {\it Phys. Rep.} \textbf{2001}, {\it 343}, 137-254
\bibitem{Scholes14}
Fassioli, F.; Dinshaw, R.; Arpin, P.\ C.; Scholes, G.\ D. Photosynthetic Light Harvesting: Excitons and Coherence. {\it J.\ R.\ Soc. Interface} \textbf{2014}, {\it 11}, 20130901-20130922
\bibitem{Reilly14}
O'Reilly, E.\ J.; Olaya-Castro, A. {\it Nat. Commun.} \textbf{2014}, {\it 5}, 3012-3021
\bibitem{Moran11}
Womick, J.\ M.; Moran, A.\ M. Vibronic Enhancement of Exciton Sizes and Energy Transport in Photosynthetic Complexes. {\it J. Phys. Chem. B} \textbf{2011}, {\it 115}, 1347-1356
\bibitem{Mancal12}
Christensson, N.; Kauffmann, H.\ F.; Pullerits, T.; Mancal, T. Origin of Long-Lived Coherences in Light-Harvesting Complexes. {\it J. Phys. Chem. B} \textbf{2012}, {\it 116}, 7449-7454
\bibitem{Mukamel13}
Dorfman, K.\ E.; Voronine, D.\ V.; Mukamel, S.; Scully, M.\ O. Photosynthetic Reaction Center as a Quantum
Heat Engine. {\it Proc. Natl. Acad. Sci. USA} \textbf{2013}, {\it 110}, 2746-2751
\bibitem{Scully97}
Scully, M.\ O.; Zubairy, M.\ S. {\it Quantum Optics} (Cambridge University Press, Cambridge, 1997)
\bibitem{Breuer02}
Breuer, H.\ -P.; Petruccione, F. {\it The Theory of Open Quantum Systems} (Oxford Univeristy Press, New York, 2002)
\bibitem{fleming10}
Ishizaki, A.; Calhoun, T.\ R.; Schlau-Cohen, G.\ S.; Fleming, G.\ R. Quantum Coherence and its Interplay with Protein Environments in Photosynthetic Electronic Energy Transfer. {\it Phys. Chem. Chem. Phys.} \textbf{2010}, {\it 12}, 7319-7337
\bibitem{Schroder06}
Schr$\ddot{\textup{o}}$der, M.; Kleinekath$\ddot{\textup{o}}$fer, U.; Schreiber, M. Calculation of Absorption Spectra for Light-Harvesting Systems using Non-Markovian Approaches as well as Modified Redfield Theory. {\it J. Chem. Phys.} \textbf{2006}, {\it 124}, 084903-084916
\bibitem{Wang08}
Wang, J.; Xu, L.; Wang, E. Potential Landscape and Flux Framework of Nonequilibrium Networks: Robustness, Dissipation, and Coherence of Biochemical Oscillations. {\it Proc. Natl. Acad. Sci. USA} \textbf{2008}, {\it 105}, 12271-12276
\bibitem{Zhedong14}
Zhang, Z.\ D.; Wang, J. Curl Flux, Coherence, and Population Landscape of Molecular Systems: Nonequilibrium Quantum Steady State, Energy (Charge) Transport, and Thermodynamics. {\it J. Chem. Phys.} \textbf{2014}, {\it 140}, 245101-245114
\bibitem{Madelung96}
Madelung, O. {\it Introduction to Solid-State Theory} (Springer-Verlag Press, Heidelberg, 1996)
\bibitem{Nazir09}
Nazir, A. Correlation-dependent Coherent to Incoherent Transition in Resonant Energy Transfer Dynamics. {\it Phys. Rev. Lett.} \textbf{2009}, {\it 103}, 146404
\bibitem{Qian82}
Qian, M.\ P.; Qian, M. {\it Z. Wahrscheinlichkeit.} \textbf{1982}, {\it 59}, 203
\bibitem{Kassal13}
Kassal, I.; Yuen-Zhou, J.; Rahimi-Keshari, S. Does Coherence Enhance Transport in Photosynthesis? {\it J. Phys. Chem. Lett.} \textbf{2013}, {\it 4}, 362-367
\bibitem{Grondelle10}
Novoderezhkin, V.\ I.; Doust, A.\ B.; Curutchet, C.; Scholes, G.\ D.; van Grondelle Excitation Dynamics in Phycoerythrin 545: Modeling of Steady-State Spectra and Transient Absorption with Modified Redfield Theory. {\it Biophys. J.} \textbf{2010}, {\it 99}, 344-352
\bibitem{Doust04}
Doust, A.\ B.; Marai, C.\ N.; Harrop, S.\ J.; Wilk, K.\ E.; Curmi, P.\ M.; Scholes, G.\ D. Developing a Structure-Function Model for the Cryptophyte Phycoerythrin 545 using Ultrahigh Resolution Crystallography and Ultrafast Laser Spectroscopy. {\it J. Mol. Biol.} \textbf{2004}, {\it 344}, 135-153
\bibitem{tiwari12}
Tiwari, V.; Peters, W.\ K.; Jonas, D.\ M. Electronic Resonance with Anticorrelated Pigment Vibrations drives Photosynthetic Energy Transfer outside the Adiabatic Framework. {\it Proc. Natl. Acad. Sci. USA} \textbf{2013}, {\it 110}, 1203-1208
\end{thebibliography}

\end{document}